\documentclass[aps,pre,reprint,superscriptaddress]{revtex4-1}

\usepackage{amssymb,amsfonts,amsmath}
\usepackage{graphicx}
\usepackage{enumerate}
\usepackage[english]{babel}
\usepackage{siunitx}
\usepackage{multirow}
\usepackage{epstopdf}

\usepackage[activate={true,nocompatibility},
            final,
            tracking=true,
            kerning=true,
            spacing=true,
            factor=1100,
            stretch=10,
            shrink=10]{microtype}

\makeatletter
  \renewcommand{\fnum@figure}{\textbf{Figure~\thefigure}}
  \makeatother

\renewcommand{\vec}[1]{\mathbf{#1}}
\newcommand{\prth}[1]{\left( #1 \right)}
\newcommand{\brcs}[1]{\left\{ #1 \right\}}
\newcommand{\sqbr}[1]{\left[ #1 \right]}
\newcommand{\avg}[1]{\left< #1 \right>}
\newcommand{\abs}[1]{\left| #1 \right|}
\newcommand{\btt}[1]{\text{\tiny{#1}}}
\newcommand{\supscr}[1]{\kern0.1em\text{#1}}
\newcommand{\kbt}{k_{\text{\tiny{B}}}T}
\newcommand{\kb}{k_{\text{\tiny{B}}}}
\newcommand{\myhati}[1]{#1\kern-0.32em\hat{\phantom{#1}}}
\newcommand{\myhatj}[1]{#1\kern-0.37em\hat{\phantom{#1}}}
\newcommand{\mbunitveci}[2]{ \hat {\boldsymbol{#1}}_{#2} }
\newcommand{\mbunitvecj}[2]{ \hat {\boldsymbol{#1}}_{\!#2} }
\newcommand{\mbunitvec}[2]{ \hat  {\boldsymbol{#1}}_{#2} }
\newcommand{\equn}{Eq.}
\newcommand{\fign}{Fig.}
\newcommand{\etal}{{\it et al.\ }}
\newcommand{\ie}{{\it i.e.\ }}
\newcommand{\eg}{{\it e.g.\ }}

\begin{document}

\title{Orientational order as the origin of the long-range hydrophobic effect}
\author{Saikat Banerjee}
\affiliation{Solid State and Structural Chemistry Unit, Indian Institute of Science, Bangalore 560012, India}
\author{Rakesh S.\ Singh}
\affiliation{Department of Chemical \& Biological Engineering, Princeton University, Princeton, NJ 08544, USA}
\author{Biman Bagchi}
\email{bbagchi@sscu.iisc.ernet.in}
\affiliation{Solid State and Structural Chemistry Unit, Indian Institute of Science, Bangalore 560012, India}
\vspace{1in}

\begin{abstract}
The long range attractive force between two hydrophobic surfaces immersed in water
is observed to decrease exponentially with their separation --
this distance-dependence of effective force is known as the hydrophobic force law (HFL).
We explore the microscopic origin of HFL by studying distance-dependent
attraction between two parallel rods immersed in 2D Mercedes Benz model of water.
This model is found to exhibit a well-defined HFL.
Although the phenomenon is conventionally explained by density-dependent theories,
we identify orientation, rather than density, as the relevant order parameter.
The range of density variation is noticeably shorter than that of orientational heterogeneity.
The latter is comparable to the observed distances of hydrophobic force.
At large separation, attraction between the rods arises primarily from a destructive interference
among the inwardly propagating oppositely oriented heterogeneity generated in water by the two rods.
As the rods are brought closer, the interference increases leading to a decrease in heterogeneity
and concomitant decrease in free energy of the system, giving rise to the effective attraction.
We notice formation of hexagonal ice-like structures at the onset of attractive region
which suggests that metastable free energy minimum may play a role in the origin of HFL.
\end{abstract}

\maketitle

\section{Introduction}

When water is confined between two large hydrophobic surfaces,
one finds the emergence of a surprisingly long ranged effective attraction between the two hydrophobes.
The attractive force grows exponentially as the distance between the surfaces is lowered. 
This exponential separation dependence of hydrophobic attraction has become widely known as the
\emph{hydrophobic force law}.

The first direct measurement of the interaction between two cylindrically curved hydrophobic surfaces 
in aqueous electrolyte solutions was performed by Israelachvili in a series of landmark experiments
using specially designed surface force measurement apparatus.~\cite{israelachvili_nature_1976, 
israelachvili_faradtrans_1978, pashley_1981, horn_jcp_1981, israelachvili_nature_1982}
These experiments observed that the attractive force between the surfaces is ``long-ranged''
-- being detectable even as far as \SIrange{80}{100}{\angstrom} separation.
It is neither sensitive to the type and concentration of the electrolyte, 
nor to the pH of the solution.~\cite{israelachvili_nature_1982}
The hydrophobic force $F$ measured in experiments can be empirically fitted to the following form
\begin{equation}
   \label{eq:hforce_law}
   \frac{F}{R}=C\exp\prth{-\frac{d}{\xi}}
\end{equation}
where $R$ is the radius of the cylindrical curve, 
$C$ is an arbitrary constant and $\xi$ is the correlation length (or, decay length).
Accurate estimate of the range of this hydrophobic force went through a period of uncertainty.
Different experimental force-measuring techniques and different methods of hydrophobization
apparently recognized substantial diversity in length scales.
There are reports of a measurable attractive force at separation as large as 
\SI{3000}{\angstrom}.~\cite{kurihara_jacs_1992,kurihara_chemlett_1990}
However, recent experiments have led to believe that the attraction at separations greater than $\SI{200}{\angstrom}$
is due to a variety of system-dependent, nonhydrophobic (or only indirectly hydrophobic-dependent) 
effects.~\cite{israelachvili_pnas_2006, israelachvili_farad_2010}
The attractive forces at separations less than $\SI{100}{\angstrom}$ represents the truly hydrophobic interaction.
The exponential decay of the attractive force between hydrophobic surfaces 
(or, the hydrophobic force law), however, remains an integral feature of all experimental results.

Despite a great deal of research over the last three decades, 
understanding of certain aspects of the origin and nature of interaction 
between hydrophobic surfaces have remained unsatisfactory. 
Theoretical studies have shown that size of the hydrophobic surface, degree of hydrophobicity 
and temperature together determine the critical separation~\cite{niharendu_jacs_2005, niharendu_jacs_2007}
as well as the rate of evaporation of confined water.~\cite{debenedetti_pnas_2012}
The phenomenon is mainly attributed to the formation of a liquid-vapor interface induced by the surface, 
leading to cavitation at some critical intersurface separation.~\cite{berne_weeks_zhou_annurev_2009}
Such cavitation is also observed near vapor-liquid coexistence for a Lennard-Jones fluid 
confined between hard walls.~\cite{attard_patey_jcp_1993}
This de-wetting induced collapse is generally described by density fields. 
In this context, Lum-Chandler-Weeks (LCW) theory~\cite{lum_chandler_weeks} has been one of the most successful approaches.
It involves the drying interface concept, and includes effects of orientation of the water molecules 
only implicitly in the Hamiltonian using prior knowledge of the radial distribution function. 
LCW theory predicts a critical intersurface separation of \SI{50}{\angstrom} for two parallel hard plates. 

The above picture basically describes water in terms of 
inward propagating density inhomogeneity originating from the two surfaces. 
However, orientational ordering induced by the surfaces is ignored.
Indeed, as early as in 1959, Kauzmann introduced the concept of ``clathrate-structure'' of water 
surrounding the hydrophobic solutes to understand the hydrophobic effect in terms of entropic stability.~\cite{kauzmann_1959} 
Formation of such structures emphasizes the role of orientational ordering in water.
Subsequent computer simulations have also found evidence of significant orientational ordering 
among water molecules around hydrophobic residues of protein molecules.~\cite{cheng_rossky_nature}
In a recent study, Berry and coworkers have discussed how oppositely directed polarization 
from water dipoles can give rise to a long range attraction.~\cite{berry_polarization}

In view of the above, the one-order-parameter (density) description of Lum \etal 
may not be adequate to describe ordering of water around a hydrophobic surface.
Hydrophobic surfaces can induce orientational inhomogeneity by disrupting 
the natural orientational correlation among water molecules.
This orientational ordering in turn may alter many fundamental properties of water, including density.
It is thus expected to play a key role in describing the metastable or unstable state, 
if any, between the two hydrophobic surfaces.
Comprehensive understanding of the long range hydrophobic attraction is a challenging task, 
particularly because of subtle nature of the hydrophobic effect itself and its interplay with an array of other forces. 

Computational approaches have proved to be very useful in this regard as one can explore the microscopic details 
and try to understand their manifestation in the macroscopic behavior.
Accurate and detailed models evidently gives improved accuracy in computer simulations.
However, simplified models often give insights that are not obtainable at relevant time scales from complex models. 
They usually capture the essence of underlying physics while being more comprehensible --
providing insights and illuminating concepts, and they do not require big computer resources.

\begin{figure}[t]
   \begin{center}
   \includegraphics[width=1.0\columnwidth]{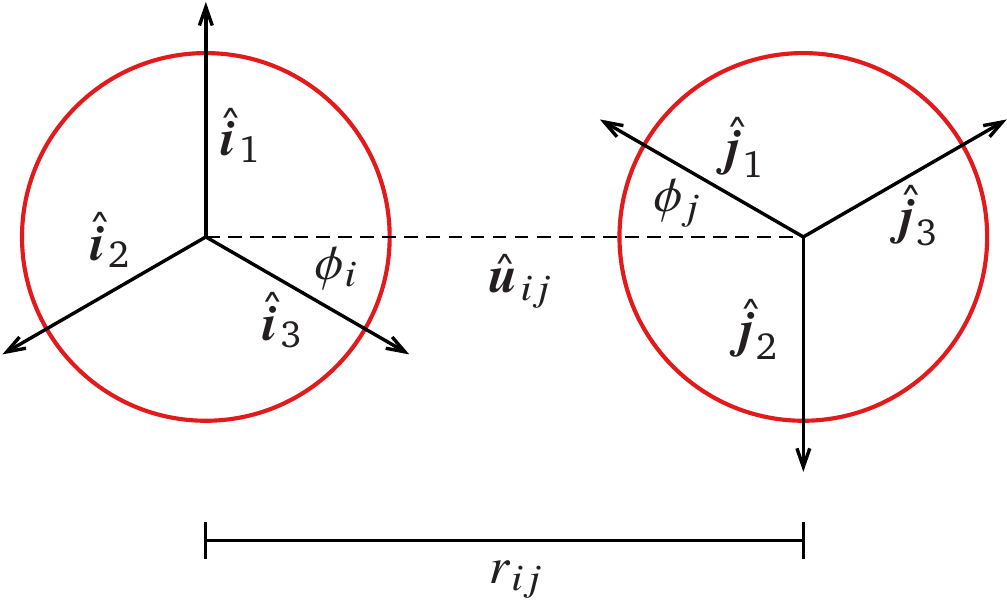}
   \end{center}
   \caption
        {  \label{fig:mb_model}
           Schematic representation of the two-dimensional Mercedes-Benz (MB) model of water.
           Two MB particles are shown, separated by a distance $r_{ij}$.
           Each particle has three H-bonding arm vectors:
           $\mbunitveci{i}{k}$ and $\mbunitvecj{j}{l}$ 
           respectively $(k, l = 1, 2, 3)$.
           The interparticle axis vector is denoted by $\mbunitvec{u}{ij}$
           and the angles that the closest arm of each particle make with the axis vector
           are labeled $\phi_i$ and $\phi_j$.
        }
\end{figure}

In the present work, we employ the Mercedes Benz (MB) model of water to study hydrophobic force law.
This model was introduced by Ben-Naim~\cite{ben-naim_jcp_1971, ben-naim_molphys_1972}, 
and later parameterized by Dill and co-workers to mimic water-like properties.~\cite{dill_mb_model_1998,
dill_jpcb_2000, dill_jcp_2000, dill_jacs_2002, dill_annrev_2005}
The merit of this model lies in the simple form of the interaction potential and the reduced dimensionality.
A schematic representation of the model is shown in \fign~\ref{fig:mb_model} 
and details are given in Materials and Methods.
Several previous studies have demonstrated that MB model generally reproduces most of the properties of water 
including the density anomaly, the minimum in isothermal compressibility as a function of temperature, 
and the thermodynamic properties of nonpolar solvation. 
This model has also been used to study the thermodynamic properties 
and structural aspects of confined water.~\cite{dill_jpcb_2006, urbic_jmolliq_2004, urbic_molphys_2014, urbic_molphys_2015}
To study the hydrophobic force, we have introduced rigid parallel hydrophobic rods made of 
non-interacting 2D Lennard-Jones (LJ) circles in MB water. 
We have performed molecular dynamics (MD) simulations in NVT ensemble 
to investigate the structure and dynamics of the system.

We find that this model system exhibits a clear attractive hydrophobic force law, with a correlation length $\xi$ 
of $\sim4\sigma$, where $\sigma$ is the diameter of the MB circle.
The force is found to decay to \SI{90}{\percent} of the contact value at around $\sim10\sigma$.
Our results and analysis provide a molecular level explanation of the origin of
long-range attractive forces between the two hydrophobic surfaces,
prior to the eventual cavitation induced collapse.
We introduced and calculated appropriate orientational order parameters.
We find that the correlation length of density heterogeneity is significantly shorter 
than that of the bond orientational order parameter, 
while the latter is comparable to the correlation length of the hydrophobic force. 
Destructive interference between orientational heterogeneities propagating inwards 
from the two surfaces lead to a lowering of imposed order, and hence lowering of free energy. 
As this destructive interference increases with decreasing separation between the two rods, 
there appears an effective attraction between the two surfaces.

\section{Hydrophobic force law in MB water}
The hydrophobic force law has been inspected in computer simulations studies of 3D water models, 
but the validity of the same in reduced dimension, such as in the MB model, has not been looked at. 
Recently, Dill and co-workers have studied the behavior of confined MB water in enclosed cavities 
using $NVT$ and $\mu VT$ Monte Carlo simulation.~\cite{dill_jpcb_2006}
The force between the walls was estimated from the oscillating density profiles and was found to be attractive.
The exponential distance-dependence and the correlation length were, however, not examined.
In this work we calculate the pressure $P_{\btt{cav}}$ in the  
confined region between the two hydrophobic rods by using the virial expression,
\begin{equation}
  \label{eq:mb_virial}
  P_{\btt{cav}} = \frac{N\kbt}{V} + \frac{\sum_{i} \vec{r}_i \cdot \vec{f}_i}{DV}
\end{equation}
where $N$ is the number of atoms in the system, $\kb$ is the Boltzmann constant,
$T$ is the temperature, $D$ is the dimensionality of the system, $V$ is the system volume,
$\vec{r}_i$ is the position vector of $i^{\supscr{th}}$ particle,
and $\vec{f}_i$ is the total internal force (arising from the interatomic interactions) 
on $i^{\supscr{th}}$ particle.
The second term in the above \equn~\ref{eq:mb_virial} is the \emph{virial}.

The pressure is found to increase exponentially with the distance of separation between the two rods.
We obtain the pressure at infinite separation $P_{\infty}$ by fitting it to the equation,
\begin{equation}
  P_{\btt{cav}} = P_{\infty} + A \exp \prth{-\frac{d}{B}}
\end{equation}
where $d$ is the inter-rod distance, $A$ and $B$ are fitting parameters.
The effective pressure on the rods is subsequently obtained as
\begin{equation}
  P = P_{\btt{cav}} - P_{\infty}
\end{equation}

\begin{figure}[t]
   \includegraphics[width=1.0\columnwidth]{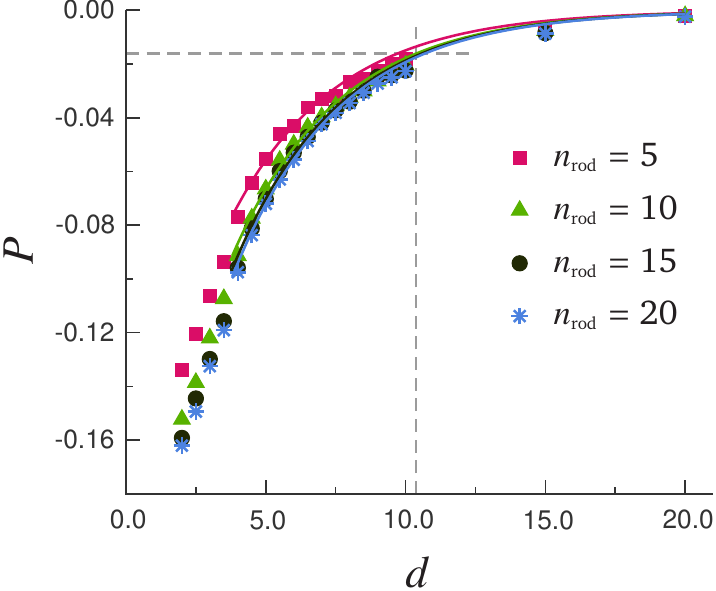}
   \caption
        {  \label{fig:mb_hforce_pressure}
           Pressure ${P}$ on each individual hydrophobic LJ rods suspended
           in MB water increases exponentially
           as the distance between the rods $(d)$ is increased.
           Simulation results are shown for different rod lengths,
           given by the number of LJ particles $(n_{\btt{rod}})$ constituting the rod.
           The solid lines are the exponential fittings [\equn~\ref{eq:hforce_pressure}]
           with global correlation length $\xi = 3.8 \pm 0.1$.
        }
\end{figure}

In \fign~\ref{fig:mb_hforce_pressure}, we show the effective pressure on the rods
as a function of inter-rod separation distance $d$.
The length of the rod is given by the number of LJ particles $(n_{\btt{rod}})$ constituting the rod.
The effective pressure shows negligible dependence on the rod length, 
and decays exponentially with inter-rod distance.
This is consistent with the experiments of Israelachvili
\etal\cite{israelachvili_faradtrans_1978, israelachvili_nature_1982}
and can be fitted to the equation
\begin{equation}
  \label{eq:hforce_pressure}
  P = C \exp \prth{- \frac{d}{\xi}}
\end{equation}
which is analogous to the hydrophobic force law \equn~\ref{eq:hforce_law}.
The solid lines in \fign~\ref{fig:mb_hforce_pressure} show the global fitting with constant $\xi$. 
The correlation length is found to be $\xi = 3.8 \pm 0.1$.
The values of the prefactor $C$ (obtained from the fitting) for the different rod lengths
are given in Table~\ref{tab:mb_prefactor}.
\begin{table}
\begin{center}
   \caption
           {Fitted values of the correlation length and prefactor of the hydrophobic force law
            (see \equn~\ref{eq:hforce_pressure}) observed between LJ rods in a system of MB particles}
   \label{tab:mb_prefactor}
   {\renewcommand{\tabcolsep}{1em}%
    \renewcommand{\arraystretch}{1.5}%
   \begin{tabular}{c | c | c} \hline \hline
      
      $n_{\btt{rod}}$   & $C$  & $\xi$\\ \hline
      $5$   & $-0.21\pm0.01$  & \multirow{4}{*}{$3.8\pm0.1$} \\
      $10$  & $-0.25\pm0.01$  & \\
      $15$  & $-0.26\pm0.01$  & \\
      $20$  & $-0.27\pm0.01$  & \\ \hline
   \end{tabular}
   }
\end{center}
\end{table}
The net force on the rods can be obtained by multiplying the effective pressure with the respective length of the rods.
Thus the correlation length $\xi$ for the effective force would not depend on the rod length, 
though the prefactors would change.

The negative pressure indicates effective attraction between the rods, induced by the intervening water.
As noted by the dashed lines in \fign~\ref{fig:mb_hforce_pressure}, 
the pressure in the confined region reaches \SI{90}{\percent} of the bulk at $d\approx10$.
Thus, there exists a tangible finite force of attraction between the two rods
even when they are separated by 10 water diameters.

Since the effect of the rod length on hydrophobic force appears negligible,
we use a particular size of the hydrophobic rod, $n_{\btt{rod}}=15$ for further analysis.

\section{Microscopic origin of the hydrophobic force: orientation and density maps}
To understand the origin of long range hydrophobic force, 
we seek to visualize the underlying control parameters -- position-dependent density and orientation -- 
of MB water in the region confined by the two parallel hydrophobic rods.
\suppressfloats[t]
\begin{figure}[t]
   {\centering
   \includegraphics[width=0.9\columnwidth]{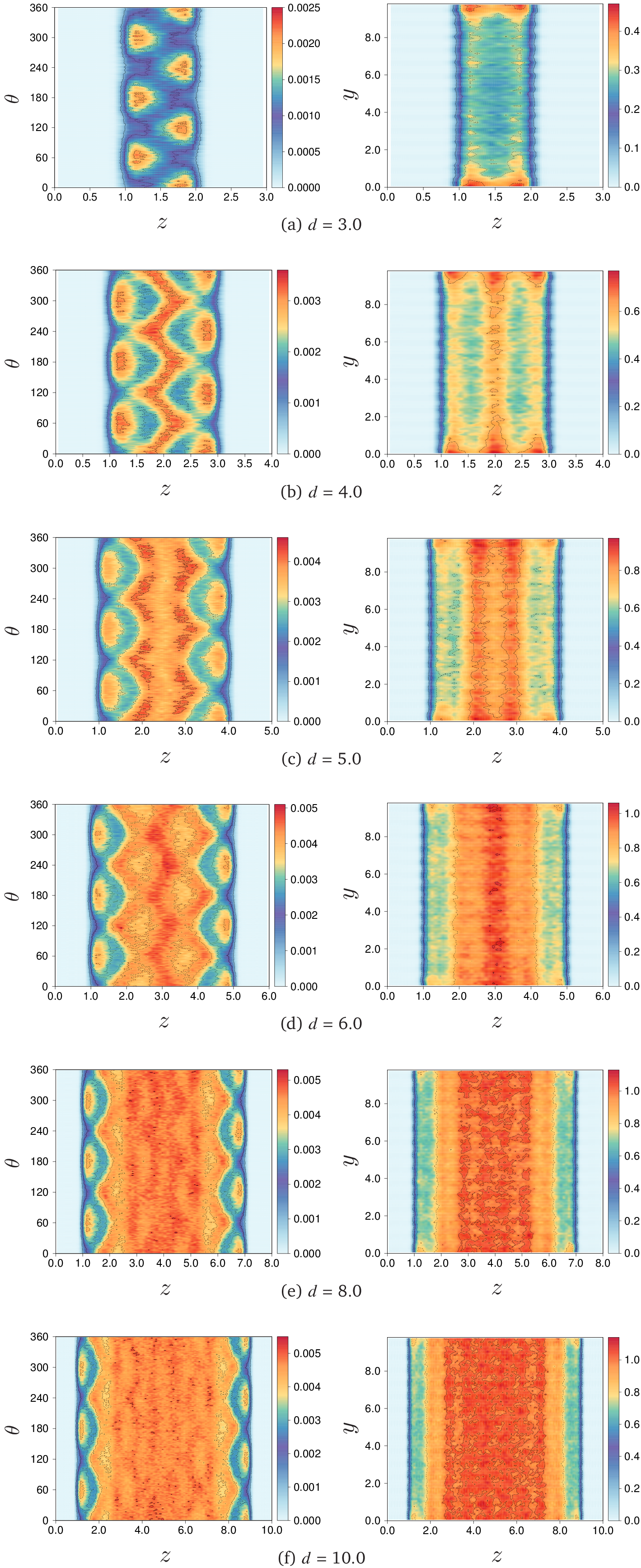}
   }
   \caption
        {  \label{fig:mb_orient_map}
           Orientation maps (left panel) and density maps (right panel)
           of MB water in the region confined between the two hydrophobic rods,
           at different inter-rod separation distance $(d)$.
           Here, $\theta$ is the angle made by the arms of the MB particles
           with the $z$-axis, \ie the axis normal to the rods.
           The left rod is at $z=0$. 
           The $y$-axis is along the length of the rods, with the bottom of the rods at $y=0$.
           The orientation maps show the probability distribution of $\theta$ along $z$.
           The density maps show the distribution of density in the $y-z$ plane.
        }
\end{figure}

We map the orientation of MB particles at different inter-rod separations 
in the left panel of \fign~\ref{fig:mb_orient_map} 
and compare with the corresponding density maps in the right panel.
The angles are measured from the $z$-axis, \ie the axis normal to the rods.
The angle zero represents an arm pointing away from the left rod.
Conversely, the angle \ang{180} represents an arm pointing away from the right rod.
The MB particles (with 3 arms) near the wall prefer a particular orientation,
with one of the arms pointing towards the wall.
Thus, particles near the left wall prefers \ang{60}, \ang{180} and \ang{300} angles,
while particles near the right wall prefers \ang{0}, \ang{120} and \ang{240}.
It means that the particles near the left wall has a dangling, unsatisfied H-bond
pointing towards the wall, and \emph{vice versa}.
The preference is due to the H-bond energy gained by the particles from 
forming ``good'' H-bonds with their remaining two arms.
Interestingly, this constraint induces an orientational pattern in the system.
Oppositely oriented patterns are created by the two hydrophobic walls. 
The inward propagation of the opposing orientational heterogeneity causes an interference
as the opposite ``waves'' meet near the center of the confined region.
This results in a frustration of orientation near the center,
and the orientational pattern survives for large inter-rod distances --
even when the rods are separated by 10 water diameters.

The density maps of MB water in the enclosed regions are
plotted alongside in the right panel of \fign~\ref{fig:mb_orient_map} for comparison.
Uniformity along the $y$-axis ensure 
that the orientational patterns of the MB particles are in dynamical equilibrium.
Visual inspection shows that oscillations in density at the center of the inter-rod region
becomes bulk-like at $d = 8.0$, while the induced orientational pattern
exists even at $d = 10.0$ and beyond.
The emergence of such exotic patterns both in orientation and density 
in such simple systems is remarkable and fascinating.
They provide a new insight to the origin of long range hydrophobic force.

\section{Distance dependence of density and orientational inhomogeneity: quantitative analysis}
One primarily employs
density-dependent descriptions, such as the LCW theory to understand hydrophobic force law.
However, as noted in the introduction,
this one-order-parameter description might not be enough to describe a complex liquid like water, 
especially due to the decrease in stability of the liquid confined within the hydrophobic surfaces.
Orientational inhomogeneity may propagate longer distances than density variation, 
as we have visualised in the preceding section.
Here, we obtain and compare the range of density and orientational inhomogeneity 
in terms of correlation lengths of the corresponding order parameters.

\subsection{Length scale and amplitude of density heterogeneity}
\begin{figure}[t]
   \includegraphics[width=1.0\columnwidth]{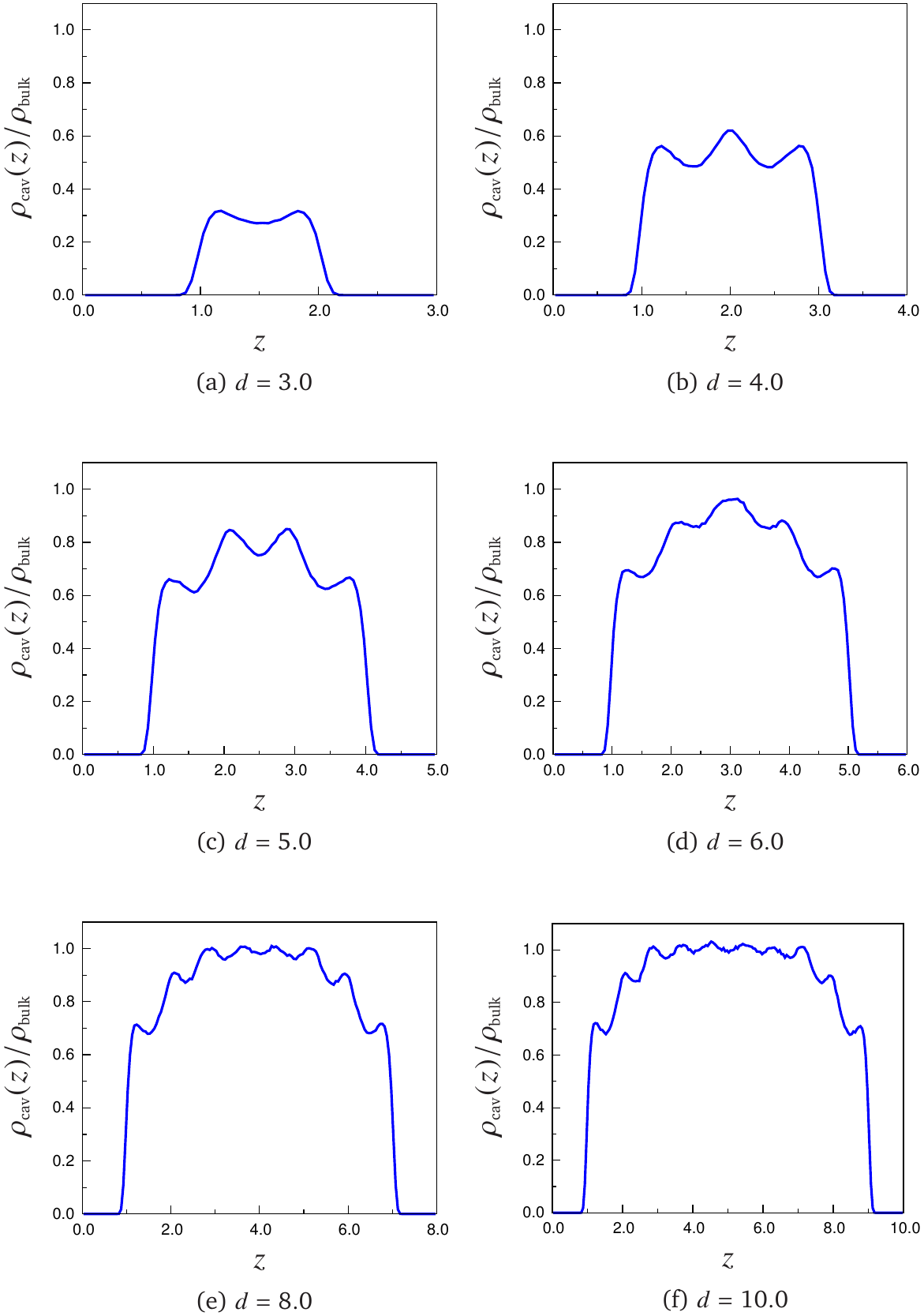}
   \caption
        {  \label{fig:mb_density_profile}
           Density profile $\rho_{\btt{cav}}(z) / \rho_{\btt{bulk}}$ of MB water 
           between the two parallel hydrophobic rods at different inter-rod separation $d$.
           The oscillating density along the $z$-direction show that the water molecules 
           are structured into multiple layers between the two rods.
           Analogous results were obtained by Dill \etal~\cite{dill_jpcb_2006} in a nanotube using $\mu VT$ simulation.
        }
\end{figure}

Density profiles of MB particles within an enclosed cavity were studied by Dill \etal~\cite{dill_jpcb_2006}.
In the present work, the particles located in between the rods are in dynamic equilibrium with the bulk.
Here, we calculate the distance-dependent density within the cavity using the same standard method,
\begin{equation}
   \rho_{\btt{cav}}(z) = \frac{\overline{N}}{Lh}
\end{equation}
where $\overline{N}$ is the average number of particles in an interval of width $h$ at distance $z$ 
from the left rod and $L$ is the length of the rods $[L=(n_{\btt{rod}}-1)\sigma_{\btt{LJ}}]$.
The density profile within the cavity is normalized by the density of the bulk $(\rho_{\btt{bulk}})$.
In \fign~\ref{fig:mb_density_profile}, we show the computed profiles of water density 
$\sqbr{\rho_{\btt{cav}}(z) / \rho_{\btt{bulk}}}$
between the two parallel rods, at different inter-rod separations $d$.
The profiles are consistent with the earlier work of Dill \etal\cite{dill_jpcb_2006}
The results show that the water molecules are structured into multiple layers between the two rods,
and hence the density oscillates along the $z$-direction.
We note that similar results were also obtained for three dimensional water models, 
studied by Choudhury and Pettitt.~\cite{niharendu_jacs_2005}

\begin{figure}[t]
   \begin{center}
   \includegraphics[width=1.0\columnwidth]{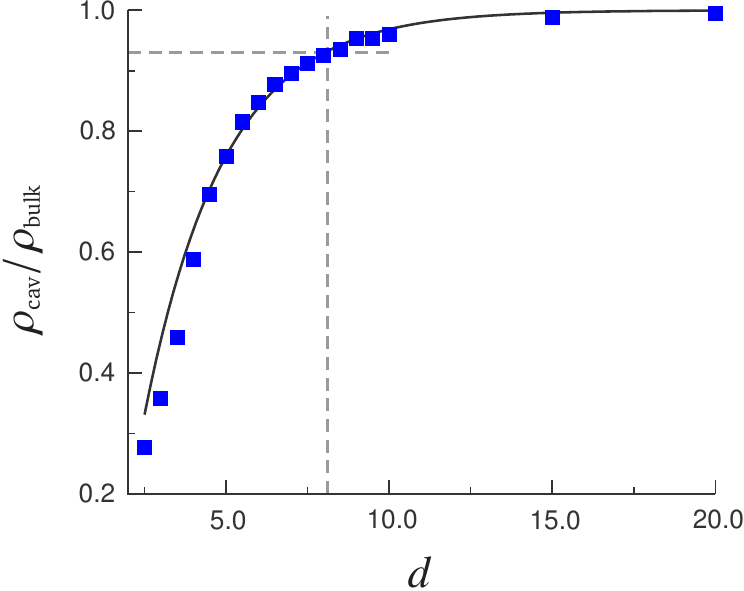}
   \end{center}
   \caption
        {  \label{fig:mb_density_order}
           Plot of density order parameter $\rho_{\btt{cav}} / \rho_{\btt{bulk}}$
           as a function of distance $d$ between the two rods.
           Here, $\rho_{\btt{cav}}$ is the average density of MB water in the confined region, 
           while $\rho_{\btt{bulk}}$ is the average bulk density of MB water.
           The solid squares are the simulation results.
           The solid line is the fitting to \equn~\ref{eq:mb_density_order}
           to obtain the correlation length, $\zeta = 2.4$.
        }
\end{figure}

The density $\rho_{\btt{cav}}$ of MB water in the region enclosed by the
two parallel hydrophobic rods gives the scalar density order parameter.
It is obtained by time averaging over the simulation trajectory,
and is normalized by the bulk density $\rho_{\btt{bulk}}$ of MB water.
In \fign~\ref{fig:mb_density_order} we show the variation of
$\rho_{\btt{cav}} / \rho_{\btt{bulk}}$ as a function
of the inter-rod separation $d$, while each rod consists of 15 LJ particles $(n_{\btt{rod}}=15)$.
The results from simulation are shown by solid squares.
The variation is found to be exponential, and can be fitted to the equation
\begin{equation}
   \label{eq:mb_density_order}
   \frac{\rho_{\btt{cav}}}{\rho_{\btt{bulk}}} = 1 + A\exp\prth{-\frac{d}{\zeta}}
\end{equation}
where $\zeta$ is the correlation length of the scalar density order parameter.
The fitting is shown by the solid line in \fign~\ref{fig:mb_density_order}.
We obtain $A=-1.86$ and $\zeta = 2.4$. 
Thus, the correlation length for density is less than the correlation length of the force between the two rods.

A useful measure to quantify the effect of the walls on the water structure
is to find the separation at which the density reaches \SI{90}{\percent} of the bulk density.
This is achieved at $d\approx8.0$ in the present case,
as shown by the dashed lines in \fign~\ref{fig:mb_density_order}.
Thus, the density of MB water in the confined region saturates to bulk density
even though there persists a significant finite force of attraction between the rods.
While the exponential decay of density may be correlated to the
exponential decay of the force at short length scales,
the origin of the attractive force beyond a certain length scale
(where cavity density saturates to bulk density)
cannot be correlated with the density order parameter.

\subsection{Length scale and amplitude of orientational heterogeneity}
In order to quantify the extent of orientational order present in the confined region,
one must define an orientational order parameter.
As we have already noted, orientations of the MB particles
are opposite to each other near the two walls.
Hence they induce opposing patterns, which propagate towards the center.
Towards this goal, we introduce the orientational order parameter $\mathrm{Q} = \avg{\cos 3 \theta}$
to quantify the orientational order propagation.
Here $\theta$ is the angle made by the arms of the MB particles,
relative to the $z$-axis, and $\avg{\ldots}$ denotes the averaging.
The value of $\avg{\cos 3 \theta}$ can range from $-1$,
when all particles are perfectly ordered with arms pointing away from the left wall,
to $+1$ when all particles are perfectly ordered with arms pointing away from the right wall.

\begin{figure}
   \begin{center}
   \includegraphics[width=1.0\columnwidth]{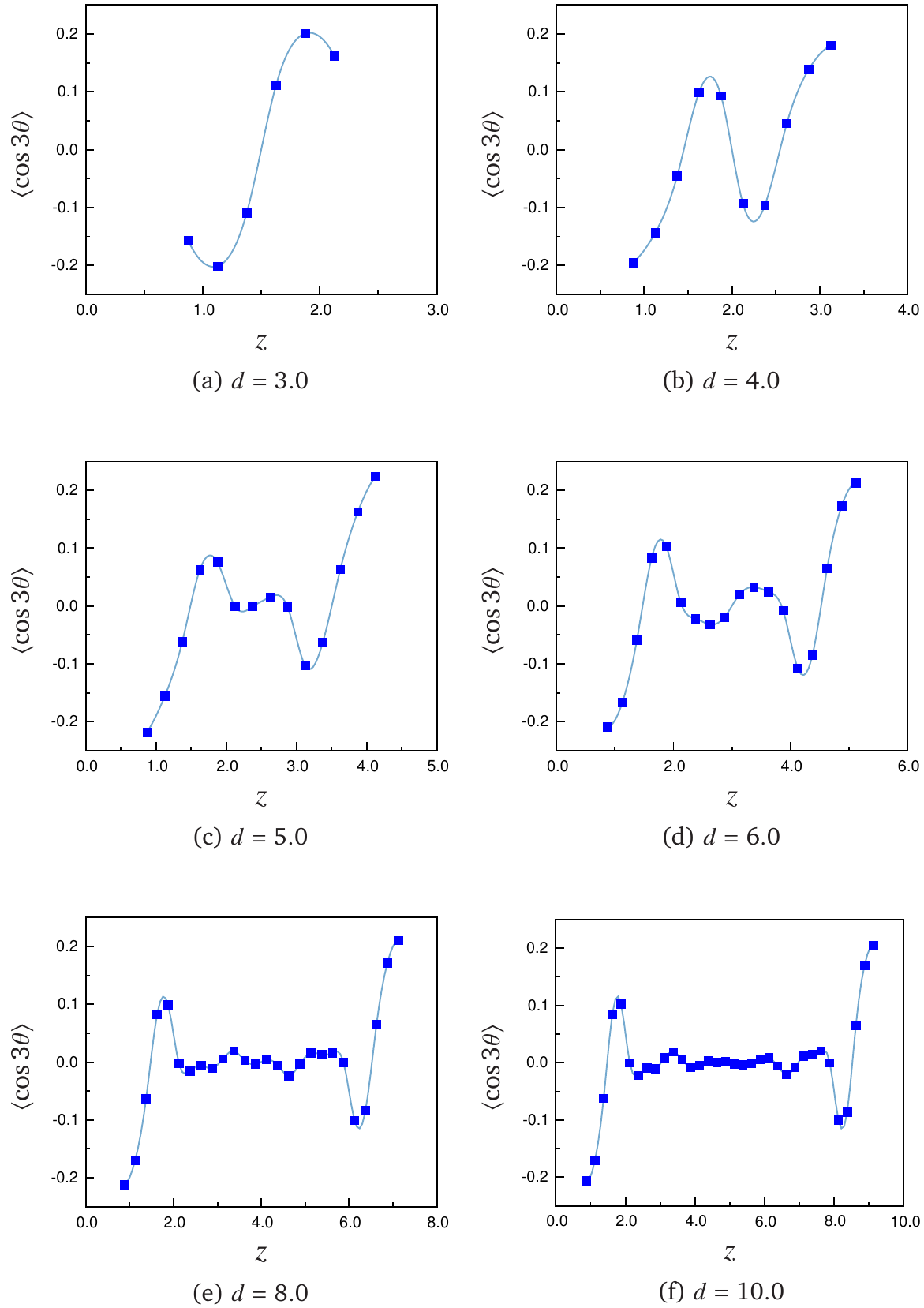}
   \end{center}
   \caption
        {  \label{fig:mb_orient_order}
           Oscillations in orientational order parameter $\avg{\cos 3 \theta}$
           along the $z$-axis of separation of the two parallel hydrophobic rods
           at different inter-rod separations $d$.
           Results from simulations are the points, whereas the solid lines are merely visual aid.
        }
\end{figure}

The order parameter $\avg{\cos 3 \theta}$ has been calculated
for MB water within small intervals of width $h$ at a distance $z$ from the left rod.
The distance-dependent propagation of $\mathrm{Q}$ is shown in \fign~\ref{fig:mb_orient_order}
at different inter-rod separations. 
We see interference of the orientational order induced by the two rods.
Strong oscillations are observed in the central region.
Adjacent positive and negative values indicate reversal of orientations,
due to interference among the inwardly propagating oppositely oriented heterogeneity.
It is interesting to note that fluctuations in $Q$
persist for longer distance than that in density profile.

\subsection{Decay of global orientational order}
A global order parameter within the cavity could be helpful for comparison
with the scalar density order parameter and formulation of a phenomenological theory.
Here we use the six-fold bond orientational order $\psi_6$, which is defined as
\begin{equation}
   \psi_6 = \frac{1}{N}\avg{\abs{\sum_{k=1}^{N}\prth{\frac{1}{N_{nb}}\sum_{j=1}^{N_{nb}}e^{i6\theta_{kj}} } }}
\end{equation}
where $N_{nb}$ is the number of nearest neighbors of a tagged particle,
$\theta_{kj}$ is the angle of the bond that the $j^{\supscr{th}}$ neighbor makes
with the tagged particle and $N$ is the total number of MB particles in the confined region.
We had earlier shown that $\psi_6$ is an effective order parameter to distinguish the
honeycomb solid and liquid state of the MB liquid~\cite{rakesh_mb_pt}.

\begin{figure}[t]
   \begin{center}
   \includegraphics[width=1.0\columnwidth]{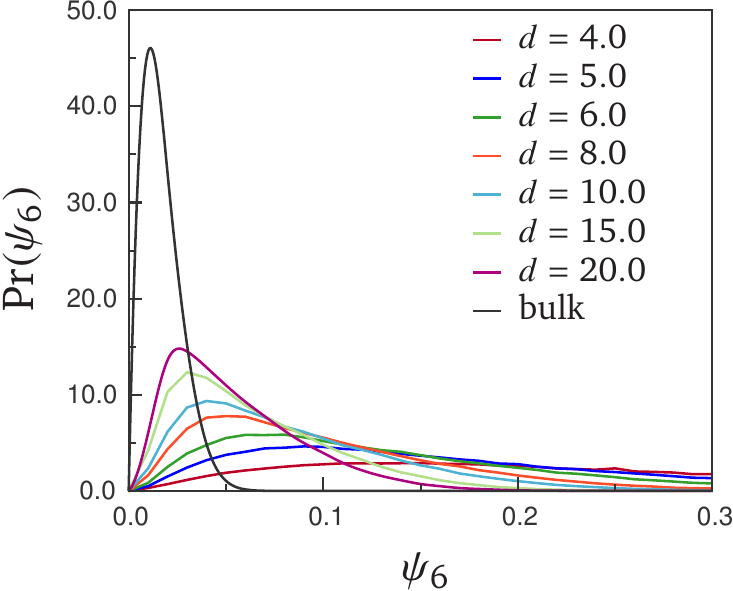}
   \end{center}
   \caption
        {  \label{fig:mb_psi6_dist}
           Probability distribution of six-fold bond orientational order $\psi_6$
           of MB liquid confined between the two hydrophobic plates.
           The distribution at different inter-rod separation $d$ are color coded, as shown in the legend.
           The long tail of the distributions of confined MB liquid, as compared to the distribution in bulk,
           indicate occurrence of ordered liquid within the cavity.
        }
\end{figure}

\begin{figure}[t]
   \begin{center}
   \includegraphics[width=1.0\columnwidth]{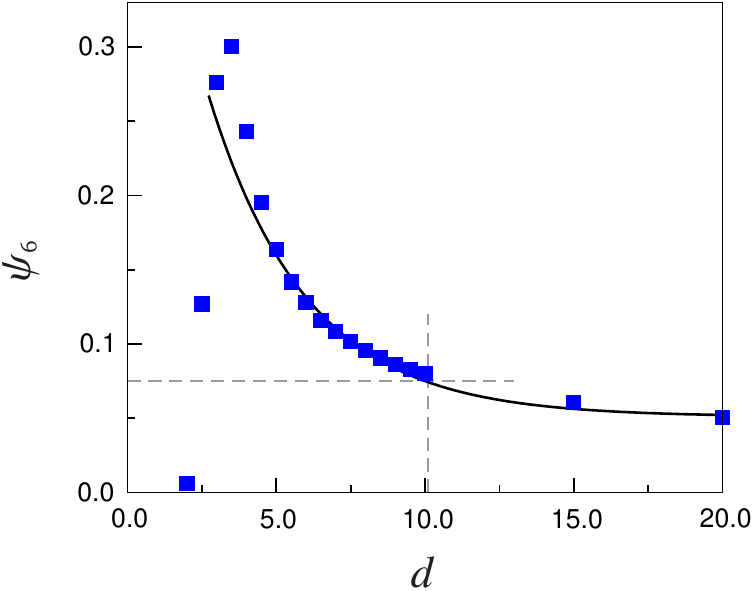}
   \end{center}
   \caption
        {  \label{fig:mb_psi6_comp}
           Average six-fold bond orientational order $\psi_6$ of MB water in the region
           between the two parallel hydrophobic rods,
           as a function of distance $d$ between the two rods.
           The solid squares are the simulation results.
           The solid line is the fitting to \equn~\ref{eq:psi6_hf_law},
           showing that the correlation length $\gamma=3.4$ propagates longer distance than density order parameter,
           and is comparable to that of the hydrophobic force (see \fign~\ref{fig:mb_hforce_pressure}).
        }
\end{figure}

In \fign~\ref{fig:mb_psi6_dist}, we show the distributions of $\psi_6$ in the confined region,
at different inter-rod separations and compare it with that of bulk MB liquid.
The long tails of the distributions reveal the ordering in the confined region.
While bulk-like behavior is approached with increasing inter-rod separation,
it is interesting to note the significant ordering in the confined region even at $d=20.0$.
In \fign~\ref{fig:mb_psi6_comp} we show the variation of $\psi_6$ as function of inter-rod separation $d$.
The results from simulation are shown by solid squares.
Initially, when there is drying in the cavity, there are not enough particles for any ordering.
At $d=3.5$ we find maximum ordering in the confined region, similar to that of an honeycomb solid, inside the cavity.
This might be due to a metastable state of the MB system (akin to low density liquid in 3D water).
The decay in orientational order is found to be exponential and can be fitted to the following form
\begin{equation}
   \label{eq:psi6_hf_law}
   \psi_6 - \psi_{\infty} = A\exp\prth{-\frac{d}{\gamma}}
\end{equation}
where $\gamma$ is the correlation length of the orientational order.
The fitting is shown as a solid line in \fign~\ref{fig:mb_psi6_comp}.
We obtain, $A=0.5$, $\gamma=3.4$ and $\psi_{\infty}=0.04$.
It is interesting to note that the correlation length for orientational order
is much longer ranged than the density order $(\zeta = 2.4)$,
and comparable to that of the hydrophobic force $(\xi=3.8)$.

\subsection{Spatio-temporal correlation of orientational relaxation}
The orientational relaxation timescale provides further insight to the dynamics of orientation.
The orientational relaxation of the MB particles is charcterized by
the orientational correlation function $C(t)$,
\begin{equation}
   C(t) = \avg{\sum_{i=0}^{N}(cos\sqbr{\theta_i(t) - \theta_i(0))}}
\end{equation}
where $\theta(t)$ is the angle made by the arms of the MB particles,
relative to the $z$-axis at time $t$,
$N$ is the number of particles in the region considered,
and $\avg{\ldots}$ denotes the ensemble average.
We calculate $C(t)$ for the MB particles
within small intervals of width $h$ at a distance $z$ from the left rod.
We have considered time average instead of the ensemble average,
since they are equivalent because of ergodicity.
The orientational relaxation is found to depend on $z$.
We fit it to a biexponential form
\begin{equation}
   C(t) = a_{1} \exp\prth{-\frac{t}{\tau_{1}}} + \prth{1-a_{1}} \exp\prth{-\frac{t}{\tau_{2}}}
\end{equation}
where the time constants for the decay of the orientational correlation
are given by $\tau_{1}$ and $\tau_{2}$,
with amplitudes $a_{1}$ and $\prth{1-a_{1}}$ respectively.
The average relaxation time $\avg{\tau}$ is subsequently obtained as
$\avg{\tau} = a_1 \tau_1 + a_2 \tau_2$ where $a_{2} = \prth{1-a_{1}}$.
\begin{figure}[t]
   \begin{center}
   \includegraphics[width=1.0\columnwidth]{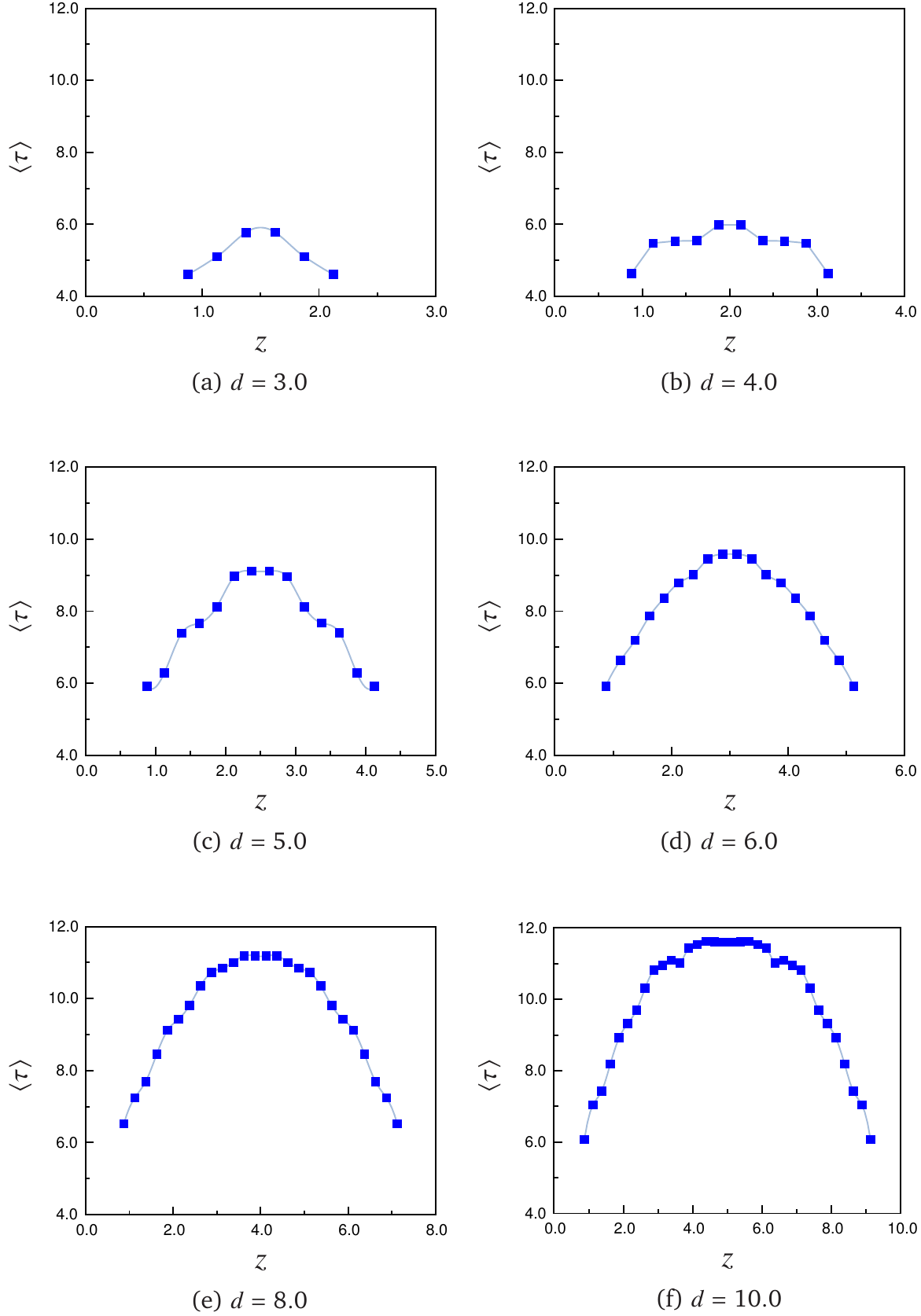}
   \end{center}
   \caption
        {  \label{fig:mb_orient_tau}
           Plots of average orientational relaxation time $\avg{\tau}$ of MB water
           along the $z$-axis of separation of the two parallel hydrophobic rods
           at different inter-rod separations $d$.
           Results from simulations are the points, whereas the solid lines are merely visual aid.
        }
\end{figure}
The variation of $\avg{\tau}$ with distance at diffferent inter-rod separations $d$ 
is shown \fign~\ref{fig:mb_orient_tau}.
The orientational relaxation is found to be faster near the hydrophobic surface,
while it reduces as we approach the bulk-like liquid near center of the confined region.
For large inter rod separations, the average relaxation time
saturates to the bulk value of $\tau_{\btt{bulk}} = 11.3$ near the center.

\section{Free energy surface for confinement induced transition: possible role of a metastable phase}
\begin{figure}[t]
   \begin{center}
   \includegraphics[width=1.0\columnwidth]{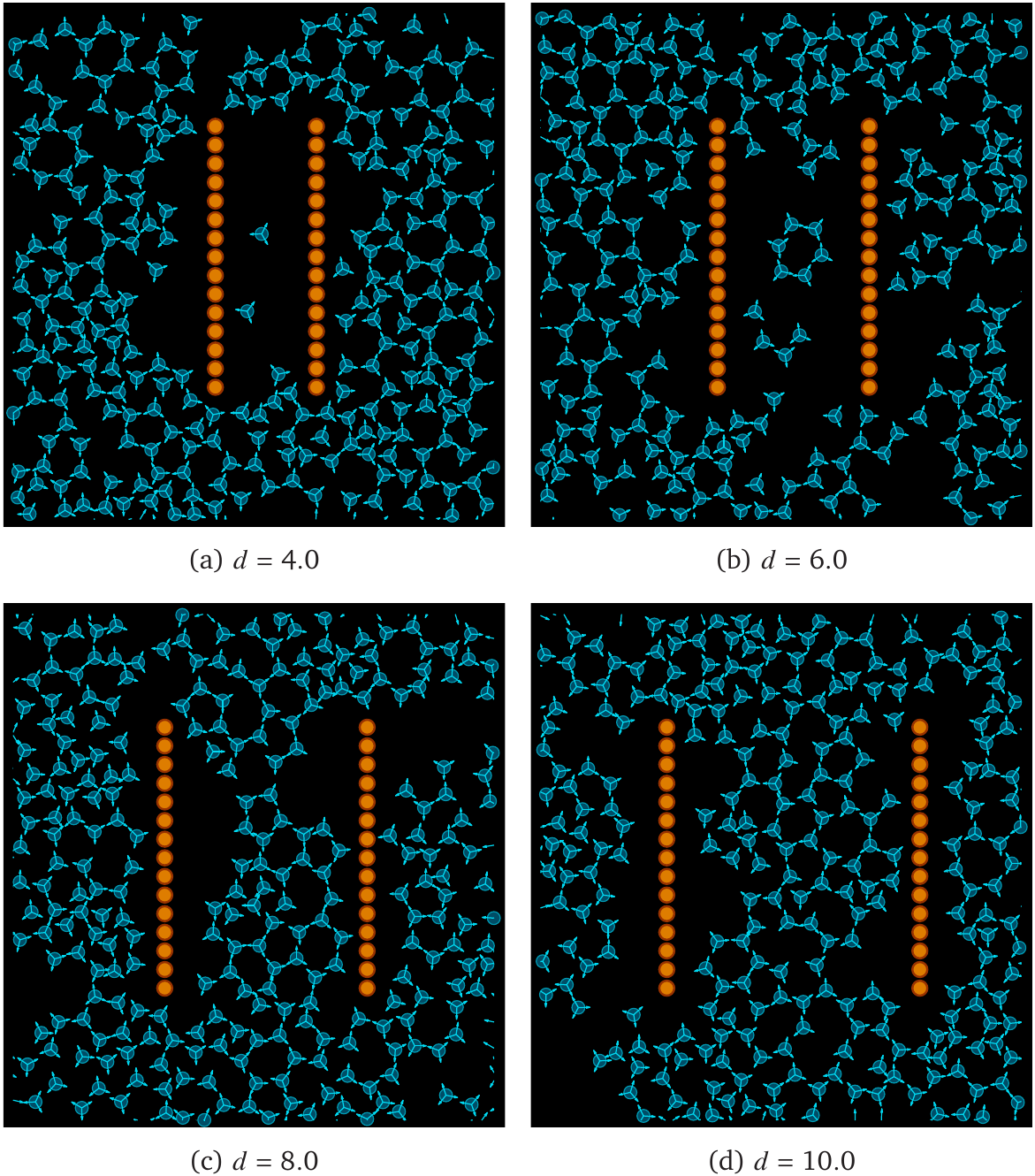}
   \end{center}
   \caption
        {  \label{fig:mb_snaps}
           Representative snapshots from the simulation, showing the region of hydrophobic confinement,
           at different inter-rod separations $d$.
           The whole box containing 2160 MB particles is not shown, instead a specific region has been focused for clarity.
        }
\end{figure}

As the density of the liquid in the confined region decreases below the bulk value
and ice-like orientational correlation increases 
(see \fign~\ref{fig:mb_orient_order}--\ref{fig:mb_orient_tau}),
additional contribution to the stability of these low density molecular arrangements may come from
the presence of nearby metastable phases.~\cite{bb_ostwald, xia_wolynes_prl}
In the present problem, the low density metastable phase is the hexagonal 2D ice (honeycomb).
For the 3D water, nearby metastable phases are ice 1h and the low density liquid(LDL)-like arrangement.
For MB particles, there is an additional metastable phase which is oblique phase.~\cite{bb_ostwald, rakesh_mb_pt, rakesh_mb_ct}
However, oblique phase is quite far off in the parameter space and we do not see any evidence of oblique phase.
However, we see strong evidence of the presence of ice-like configurations (see \fign~\ref{fig:mb_snaps}).
As discussed elsewhere~\cite{bb_ostwald},
metastable minima in free energy can lower the surface energy
and thereby play an important but ``hidden'' role in the stability of the confined liquid,
and hence provide additional attraction at some separation.

As mentioned earlier, existing theories of hydrophobic force are based 
primarily on number density as the sole order parameter 
that varies across the system as we travel from one surface to another. 
As a result, they can miss important physical constraints coming explicitly from orientation of the water molecules,
such as the clathrate-like structures with pronounced orientational ordering.~\cite{kauzmann_1959,
head-gordon_pnas_1995,cheng_rossky_nature}
It becomes particularly important when the separation is somewhat large, 
may be of the order of 8--10 molecular diameters wide in the present MB system. 
At such separations, the density inhomogeneity become rather small 
but orientational heterogeneity may still be significant, as we indeed find in our simulations. 
This suggests that orientational heterogeneity is important and needs to be taken into account.
The crucial role of orientation was earlier pointed out by Tanaka~\cite{tanaka_bo_review}, 
in case of crystallization, quasi-crystal formation, glass transition, etc. in 3D liquids.
In fact, for water-like systems, orientation could be more important of the two
as density is slaved to orientation imposed by H-bonding.

Physical picture becomes more interesting as we move to larger $d$ due to enhanced population of 6-membered rings.
In \fign~\ref{fig:mb_snaps}, we show some representative snapshots of the simulations
at different inter-rod separations.
We find an increase in the presence of hexagonal ice-like rings
at intermediate separation distances $(d=8-12)$ between the rods.
As already discussed, this corroborates with the argument that the system 
experiences the metastable free energy minimum due to the ice phase.
Clearly, such a minimum can help lowering surface free energy if the system permits
orientational arrangement close to ice.
Cooperativity of water molecules induced by the hydrophobic surface 
has been observed in 3D water models as well.~\cite{head-gordon_pnas_1995,cheng_rossky_nature}

\begin{figure}[t]
   \begin{center}
   \includegraphics[width=1.0\columnwidth]{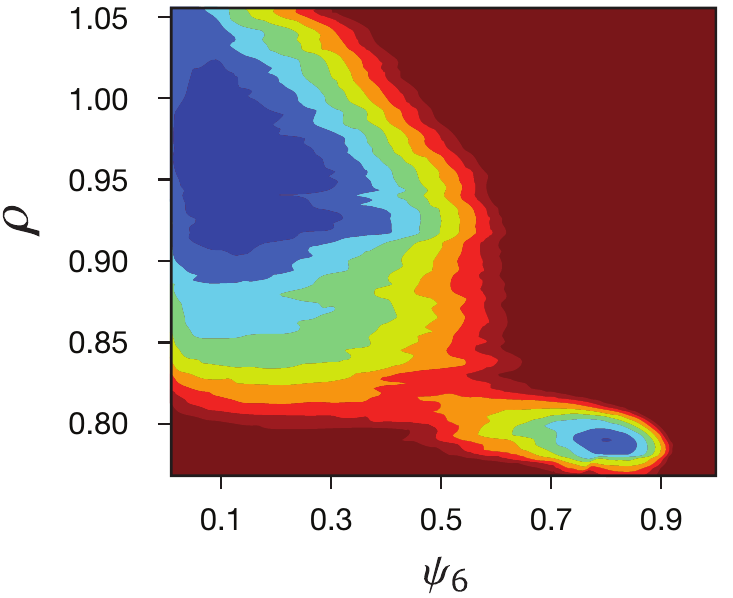}
   \end{center}
   \caption
        {  \label{fig:mb_fes}
           Free energy surface of the Mercedes Benz system at $P=0.19$ and $T=0.15$.
           Here, $\rho$ is the density of the liquid and $\psi_6$ is the six-fold bond orientational order.
           It shows the liquid phase at high $\rho$ and low $\psi_6$, 
           while the honeycomb phase (akin to low density ice) appears at lower $\rho$ and higher $\psi_6$.
        }
\end{figure}

In \fign~\ref{fig:mb_fes} we show the calculated free energy surface corresponding to the melting of MB system 
at $P=0.19$ and $T=0.15$.
The minimum at high $\rho$ and low $\psi_6$ corresponds to the liquid phase, 
whereas the honeycomb phase (akin to low density ice) creates another minimum at lower $\rho$ and higher $\psi_6$.
The increase in orientational order and decrease in density, as observed in our simulations 
and visualized in the snaps (\fign~\ref{fig:mb_snaps}) may correspond to this metastable minimum.
While LCW theory considers the cavitation and gas-liquid transition for collapse of large hydrophobic surfaces,
\emph{this metastable minimum should be considered for longer ranged attraction.}

\begin{figure}[t]
   \begin{center}
   \includegraphics[width=1.0\columnwidth]{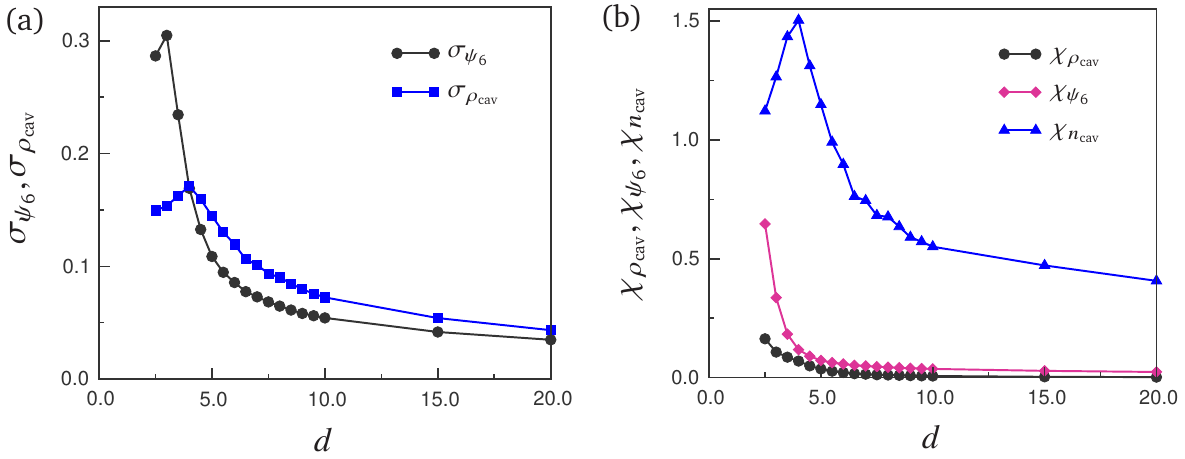}
   \end{center}
   \caption
        {  \label{fig:op_profiles}
           Plots of order parameter fluctuations and susceptibilities.
           (a) Fluctuations in orientational order parameter $(\sigma_{\psi_6})$
               and cavity density $(\sigma_{\rho_{\btt{cav}}})$.
           (b) Compressibility $(\chi_{n_{\btt{cav}}})$ of the MB liquid in the confined region,
               and comparison with susceptibilities of orientation $(\chi_{\psi_6})$
               as well as cavity density $(\chi_{\rho_{\btt{cav}}})$.
           The susceptibilities are defined in \equn~\ref{eq:compressibility}.
        }
\end{figure}

The above conclusions are substantiated by the sharp rise in fluctuation of density and orientational order
as shown in \fign~\ref{fig:op_profiles}.
We show the variance of density and orientation order parameters
at different inter-separation distances in \fign~\ref{fig:op_profiles}a.
The crossover at $d=4.0$ suggests increased fluctuation and a corresponding free energy minimum,
which might be due to metastable hexagonal phase.
Another interesting way to explore the underlying free energy is to study the susceptibility
for the different properties, defined as,
\begin{equation}
   \label{eq:compressibility}
   \chi_A = \frac{\avg{A^2} - \avg{A}^2}{\avg{A}} = \frac{\sigma_A^2}{\avg{A}}
\end{equation}
where $A$ is the property under consideration.
When we consider the number of MB particles in the cavity $(n_{\btt{cav}})$,
then $\chi_{n_{\btt{cav}}}$ gives the compressibility of the system.
As shown in \fign~\ref{fig:op_profiles}b, the compressibility goes through a maximum
at $d=4.0$, suggesting a metastable minimum.

A phenomenological description of hydrophobic force should therefore have at least two order parameters,
namely density and orientational order (which could be bond-orientational order depending on the system). 
In this case of MB liquid, density between the two phases (liquid and honeycomb) differs by $\sim$ \SI{20}{\percent},
whereas the bond orientational order differs by $\sim$ \SI{75}{\percent}.
A Ginzburg--Landau free energy functional can be written in terms of two order parameters,
one conserved (number density) and one non-conserved (orientational density),
to account for the non-local inhomogeneity,
\begin{equation}
   \label{eq:ginzburg-landau}
   \vec{F}_{\btt{tot}}\brcs{\psi_6,\rho} = \vec{F}_{\psi_6}   \brcs{\psi_6}
                                          +\vec{F}_{\rho}     \brcs{\rho}
                                          +\vec{F}_{\btt{int}}\brcs{\psi_6,\rho}
\end{equation}
The first two terms on the right hand side correspond to each order parameter, 
and the third term considers the interaction between them.

The free energy per particle of the MB system should be maximum near the walls due to higher ordering,
and should gradually reduce towards the bulk.
One has to consider that the ordering is induced on both sides of the rod,
so that the excess total free energy of the system for infinitely separated rods is 
$4\vec{F}_{\mathrm{Q}}^{\btt{max}}$, where  $\vec{F}_{\mathrm{Q}}^{\btt{max}}$ 
is the maximum free energy obtained by allowing the order to propagate on any one side up to infinity.
As the two rods are brought closer, the opposing orders interfere,
creating a frustration and reduction of order at the center (see \fign~\ref{fig:mb_orient_map}).
Thus, the total free energy of the system decreases up to the extreme limit of 
$2\vec{F}_{\mathrm{Q}}^{\btt{max}}$ when the two rods collapse.
Asymptotically, the total change in free energy in going from infinite separation 
to collapsed state is $-2\vec{F}_{\mathrm{Q}}^{\btt{max}}$.
This explains semi-quantitatively the attraction between the two rods at longer length scales.

\section{Concluding remarks}
Long range hydrophobic attraction plays a key role in several biological self-organizing processes,
such as protein folding, assembly of membrane structures, ligand binding to hydrophobic patches on target proteins,
micellar aggregation, lipid bilayer formation, etc.~\cite{bryngelson_wolynes_pnas_1987,
bryngelson_wolynes_jpc_1989,chan_dill_1997,southall_dill_haymet_jpcb_2002}
It is also crucial to phenomena such as mineral flotation, wetting, coagulation,
and many processes that involve surfactant aggregation, \eg detergency.
Deciphering the hydrophobic interactions can also have profound implications
for drug design and delivery to specific target proteins in a cell.~\cite{plumridge_jphp_drug}

The present work demonstrates that a significant, attractive force can exist 
between two hydrophobic rods in a simple 2D mmolecular liquid.
The system exhibits an exponential increase in the attractive force
as the two rods are brought closer together. 
In addition, even the estimate of the range of the force is in rough agreement
with experimentally observed range.

Although the MB model does not possess the complexity of interactions experienced by real water molecules, 
it mimics many of the important anomalies and properties of water. 
The simplicity of the present model and the ease of visualization
can therefore be harnessed to gain important insights needed to understand the puzzling 
(and often controversial) aspects of water. 
These are sometimes obscured by the complexity of the 3D models. 

Microscopic analysis of the structure and orientation of the system reveals several important issues. 
The length scale of density fluctuations is found to be shorter in range 
than the length scale of hydrophobic force between the rods.
In contrast, orientational heterogeneity propagates to longer lengths
with a correlation length that compares well with that of the hydrophobic force law. 
Hydrophobic force law is found to be a consequence of the minimum frustration of the extended H-bond network 
-- the frustration being imposed by the hydrophobic surface.
Destructive interference between the orientational heterogeneity induced by the two surfaces 
creates a metastable ice-like phase in the central region, thereby lowering the free energy of the system.

With the insight gained from the MB model, 
it would be interesting to see an extension of this work to 3D water models 
with specific attention to orientation of the water molecules, 
which has largely been overlooked in past investigations. 
A detailed theoretical analysis with the orientational order parameter may be rewarding, 
and could complement our present understanding. 
On a slightly different note, pattern formation in nature 
has been a fascinating and demanding field of research 
and the interfering orientational patterns appearing in this system (see \fign~\ref{fig:mb_orient_map})
could be an excellent model to study.
Role of orientation in the hydrophobic force law indicates 
that such interference as observed here may play an important role in directing protein association and self-assembly
where attraction between extended hydrophobic surfaces is believed to be involved.

{\small
\section*{Materials and Methods}
\subsection{Mercedes Benz model}
In the Mercedes-Benz (MB) model, water molecules are represented
as two dimensional disks with three symmetrically arranged arms,
separated by an angle of \ang{120}, as in the Mercedes Benz logo.
Each molecule interacts with other through
an isotropic Lennard-Jones (LJ) and an anisotropic hydrogen-bond (HB) term
in a completely separable form as
\begin{equation}
  U\prth{\vec{X}_i, \vec{X}_j} = U_{\btt{LJ}}\prth{r_{ij}} + U_{\btt{HB}}\prth{\vec{X}_i, \vec{X}_j}
\end{equation}
where $\vec{X}_i$ denotes vectors representing both the coordinates and orientation
of the $i^{\supscr{th}}$ particle and $r_{ij}$ the distance between the centers of two particles.
The notations are summarized in \fign~\ref{fig:mb_model}.
The LJ term $U_{\btt{LJ}}$ is defined as
\begin{equation}
    U_{\btt{LJ}}\prth{r_{ij}} = 4\epsilon_{\btt{LJ}} \sqbr{\prth{\displaystyle\frac{\sigma_{\btt{LJ}}}{r_{ij}}}^{12}
                                                          -\prth{\displaystyle\frac{\sigma_{\btt{LJ}}}{r_{ij}}}^{6 }
                                                          }
\end{equation}
where $\epsilon_{\btt{LJ}}$ and $\sigma_{\btt{LJ}}$
are the well-depth and diameter for the isotropic LJ interaction.
The anisotropic HB part of potential, $U_{\btt{HB}}$ is defined as
\begin{eqnarray}
    && U_{\btt{HB}}\prth{\vec{X}_i, \vec{X}_j}  = \nonumber\\
    &&                                 \epsilon_{\btt{HB}}G(r_{ij}-r_{\btt{HB}})
                                        \sum_{k,l=1}^{3} G\prth{\mbunitveci{i}{k}\cdot\mbunitvec{u}{ij} - 1}
                                                         G\prth{\mbunitvecj{j}{l}\cdot\mbunitvec{u}{ij} + 1} \nonumber\\
    &&
\end{eqnarray}
where $G(x) = \exp(-x^2/2\sigma^2)$.
This form of the anisotropic potential ensures that neighboring water molecules
form explicit H-bonds (have favorable energy)
when the arm of one molecule aligns with the arm of another --
the strongest bond occurs when they are perfectly collinear
and are opposite in direction.
The unit vector $\mbunitveci{i}{k}$ represents the $k^{\supscr{th}}$ arm of the
$i^{\supscr{th}}$ particle $(k = 1, 2, 3)$ and $\mbunitvec{u}{ij}$ is the unit vector
joining the center of particle $i$ to the center of particle $j$.
We have used the optimal parameters of the MB model that are known to reproduce the properties of water, 
$\epsilon_{\btt{HB}}=-1.0$, $r_{\btt{HB}}=1.0$, $\epsilon_{\btt{LJ}}=0.1$, $\sigma_{\btt{LJ}}=0.7$
and $\sigma_{\btt{HB}}=0.085$.

\subsection{Simulation details}
In this study, we use MB model of water.
The rigid hydrophobic rods are made up of non-interacting 2D Lennard-Jones (LJ) particles.
The LJ particles were separated by a distance of $0.7\sigma_{\btt{HB}}$.
For the interaction between the LJ particles and MB particles,
we considered the same parameters as used by
Debenedetti \etal~\cite{debenedetti_pnas_2012} in their 3D simulation,
but in reduced units, $\epsilon_{\btt{LJ-MB}}=\num{0.006}$ and $\sigma_{\btt{LJ-MB}}=\num{1.2}$.
We incorporated the 2D MB force field in LAMMPS for carrying out Molecular Dynamics (MD) simulation.
The forces and torques were calculated using similar techniques as used earlier~\cite{3D_MB_MD} for the 3D model.

We generate an initial configuration of \num{2160} MB particles in a honeycomb lattice.
We perform equilibration at temperature, $T = 0.2$
and pressure $P = 0.19$ to obtain a liquid configuration.
We insert two rods at required distance by replacing any MB particles, if within contact distance. 
The system was equilibrated for $10^5$ steps at constant temperature and volume ($NVT$ ensemble),
with each time step $\tau = \num{0.007}$. 
The production run was carried out for $10^7$ steps at constant temperature and volume.
We used Nose-Hoover thermostat to keep the temperature bath at $T = 0.2$.

\section*{Acknowledgements}
We thank Prof. P. G. Wolynes for his valuable comments on the manuscript
and insight to the problem.
BB thanks Prof. I. Ohmine for a useful discussion. 
This work was supported in parts by grants from BRNS (Mumbai) and DST (Delhi).
We thank DST for a JC Bose Fellowship (to BB).

}

\end{document}